\documentclass[useAMS,usenatbib]{mn2e}
\usepackage{aas_macros}
\usepackage{url,ulem,times,graphicx,amsmath,amsfonts,amssymb,color,epsfig,epstopdf}
\usepackage[a4paper,centering, totalwidth=520pt, totalheight=700pt]{geometry}
\usepackage{epsf}
\usepackage{bm}
\usepackage{color}
\usepackage[T1]{fontenc}
\usepackage{multirow,textcomp}
\def\eone{${\bf e}_{1}\,$}
\def\etwo{${\bf e}_{2}\,$}
\def\ethree{${\bf e}_{3}\,$}

\title[The alignment of galaxy spin with the shear field in observations]
  {The alignment of galaxy spin with the shear field in observations}
\author[Pahwa et al]
{Isha Pahwa$^1$, Noam I. Libeskind$^1$, Elmo Tempel$^2$, Yehuda Hoffman$^3$, R. Brent Tully$^4$, \newauthor H$\acute{{\rm e}}$l$\grave{\rm e}$ne M. Courtois$^5$, Stefan Gottl\"{o}ber$^1$, Matthias Steinmetz$^1$ and Jenny G. Sorce$^1$\\
\vspace*{0.1cm}\\
$^1$Leibniz-Institut f\"ur Astrophysik Potsdam (AIP), An der Sternwarte 16, D-14482 Potsdam, Germany\\
$^{2}$Tartu Observatory, Observatooriumi~1, 61602 T\~oravere, Estonia\\
$^3$Racah Institute of Physics, Hebrew University, Jerusalem 91904, Israel\\
$^4$Institute for Astronomy, University of Hawaii, 2680 Woodlawn Drive, HI 96822, USA\\
$^5$Universit$\acute{{\rm e}}$ Lyon 1, CNRS/IN2P3, Institut de Physique Nucl$\acute{\rm e}$aire, Lyon, France
}
\date{Accepted --- . Received ---; in original form ---}

\pagerange{\pageref{firstpage}--\pageref{lastpage}} \pubyear{2015}

\def\LaTeX{L\kern-.36em\raise.3ex\hbox{a}\kern-.15em
    T\kern-.1667em\lower.7ex\hbox{E}\kern-.125emX}

\begin{document}

\label{firstpage}

\maketitle

\begin{abstract}
Tidal torque theory suggests that galaxies gain angular momentum in the linear stage of structure formation. Such a theory predicts alignments between the spin of haloes and tidal shear field. However, non-linear evolution and angular momentum acquisition may alter this prediction significantly. In this paper, we use a reconstruction of the cosmic shear field from observed peculiar velocities combined with spin axes extracted from galaxies within $115\, \mathrm{Mpc} $  ($\sim8000 \, {\mathrm {km}}{\mathrm s}^{-1}$) from 2MRS catalog, to test whether or not galaxies appear aligned with principal axes of shear field. Although linear reconstructions of the tidal field have looked at similar issues, this is the first such study to examine galaxy alignments with velocity-shear field. Ellipticals in the 2MRS sample, show a statistically significant alignment with two of the principal axes of the shear field. In general, elliptical galaxies have their short axis aligned with the axis of greatest compression and perpendicular to the axis of slowest compression. Spiral galaxies show no signal. Such an alignment is significantly strengthened when considering only those galaxies that are used in velocity field reconstruction. When examining such a subsample, a weak alignment with the axis of greatest compression emerges for spiral galaxies as well. This result indicates that although velocity field reconstructions still rely on fairly noisy and sparse data, the underlying alignment with shear field is strong enough to be visible even when small numbers of galaxies are considered -  especially if those galaxies are used as constraints in the reconstruction.
\end{abstract}


\section{Introduction}
The origin of galactic spin is a well posed problem in structure formation. It is now accepted that tidal torques from 
nearby structures can generate angular momentum in collapsing clouds \citep{stromberg1934,hoyle1951,peebles1969,doroshkevich1970,white1984},
at least during the linear phase of structure formation. Such a theory predicts correlations between the tidal field and 
galaxy spin, correlations that are imparted in linear regime. Specifically, the theory predicts that the angular momentum 
generated is $L_{i}=\epsilon_{ijk}I_{jl}T_{lk}$, where $I_{jl}$ is the inertia tensor, $T_{lk}$ is the tidal field and $i,j,k=x,y,z$ are the spatial coordinates. 
In the frame of the tidal tensor, $L_{1}\propto(\lambda_{2}-\lambda_{3})I_{23};~L_{2}\propto(\lambda_{1} -\lambda_{3})I_{13};~L_{3}\propto(\lambda_{2}-\lambda_{1})I_{21}$, 
where $\lambda_{i}$'s are the ordered eigenvalues of the tidal shear tensor \citep[see][]{lee+2001}. Since $\lambda_{1}-\lambda_{3}$ is by definition 
the largest, $L_{2}$ is the greatest. Therefore, in the absence of any non-linear evolution, tidally generated angular momentum should be well aligned with the axis of intermediate collapse \citep{lee+2007}. In general, such an alignment has been only weakly confirmed -- non-linear structure formation can significantly alter these alignments \citep{lee+2007} and impart torques that work against those engendered in the linear regime. However, measuring such alignments is fairly challenging. The study of these alignments is essential for the next generation weak lensing measurements and cosmological constraints \citep{heavens+2000,croft+2000,hirata+2004,scodis+2014}.

In order to test these theoretical predictions two important, albeit difficult, measurements have to be made: a principal 
component analysis of the tidal field or Large Scale Structure (LSS) and a measurement of a galaxy's spin axis. The latter 
is very hard to do when all galactic morphologies are considered (since, e.g., non-rotating early type galaxies, have 
poorly defined spin axes). Furthermore, line of sight degeneracies with respect to the inclination and lack of the knowledge regarding how a galaxy's shape 
and spin are aligned also hamper this aspect of the problem. But progress has been made if only thin, edge-on disc galaxies 
are used (e.g Navarro et al 2004). On the other hand, the LSS of the galaxy distribution -- also known as the cosmic web 
\citep{joeveer+1978,bond1996} can be quantified using a variety of techniques given its highly contrasted, multi-scale, multi-dimensional web-like nature. 

Methods to quantify the cosmic web abound  \citep{Stoica+2005, Shen+2006, Aragon-Calvo+2007, Hahn+2007, Sousbie+2008, Forero-Romero+2009, hoffman+2012, Libeskind+2012, noam+2013, tempel+2014filament}  depending on the data and problem at hand. In general, the methods based on simulations rely on a more complete knowledge of the cosmic fluid than what is usually available in observations. Nevertheless, studies that compare similar methods across simulations and observations have found some degree of agreement \citep{tempel+2014align, tempel+2015, libeskind+2015}. For the 
purpose of checking galaxy alignments, a measure of the tidal shear field \citep[e.g.][]{Hahn+2007,Forero-Romero+2009,hoffman+2012} is needed.

When combined with knowledge of galaxy orientations, tidal torque theory (TTT) can be empirically tested. A number of 
studies \citep[e.g.][]{kashikawa+1992, navarro2004, trujillo+2006}, have for the most part focused on the alignment of 
galactic spin with one specific cosmic web environment, asking, for example, ``how do galaxies in filaments spin?''  
\cite{navarro2004} found that (spiral) galaxies in the supergalactic plane tend to spin with their spin axis in the supergalactic 
plane. \cite{trujillo+2006} also found that spin axes of spiral galaxies are preferentially in the plane of the walls of cosmic voids, although this signal appears to be disputed by \cite{slosar+2009} who found no alignment between spiral galaxies and void walls. If sitting in a cosmic filament, \cite{tempel+2013} and \cite{tempel+2013b} used the SDSS to show that the short axes of elliptical galaxies tend to be perpendicular to the filament axis.  On the other hand, the spin axes of bright spirals have a weak tendency to be aligned along the filament's spine. This spin-flip was found in numerical simulations as well \citep{Aragon-Calvo+2007,Codis+2012,noam+2013} and found to be mass and environment dependent. Below a certain mass, haloes spin parallel to filaments, above a certain mass they spin perpendicular to them. This flipping has been interpreted as being due to different formation mechanism \citep{welker+2014}. Small haloes form through diffuse accretion and not major mergers - as such they are wound up at early times and thus spin parallel to the filament in which they reside. More massive haloes form by mergers that come in along the filament and hence spin perpendicular to it. Such a hypothesis essentially suggests that the spin alignment reflects the merger history of the halo/galaxy. A number of recent reviews \citep{joachimi+2015, kirk+2015, kiessling+2015} summarize the current status of spin alignments with the cosmic web both in observations and simulations.

Where most of these studies agree, is that  although statistically significant, the alignment is weak, occurring at the most 10-20\% of the time \citep{tempel+2013, tempel+2013b}. Note that none of the above studies  \citep[except][]{lee+2007} found a close alignment with the intermediate axis of collapse, as suggested by linear TTT.  The one study that did so, \cite{lee+2007}, found a very weak alignment with  the intermediate axis, entirely driven by galaxies in dense environments, where the (linear) reconstructed tidal field is less prone to inaccuracies. Indeed, in most of the studies mentioned above, the lack of signal could be due to the fact that the intermediate axis was not measured or known.

Two issues remain open: are alignments environmentally dependent? Do they occur more or less frequently in, say, a cosmic sheet, filament or void?  Secondly, do alignments between galaxy spin and the LSS reflect what we naively expect from tidal torque theory, or if not, can they provide insight on non-linear structure formation?

In the present study, we attempt to address these questions by examining the alignment of galaxy spin with the principal axes of the cosmic web, reconstructed using the Cosmicflows-2 \citep{tully+2013} - survey of peculiar velocities. We test this alignment for galaxies in one of the biggest data sets for the local Universe - the 2MASS galaxy redshift survey.

\section{DATA and Method}
\label{sec:data}
Here we describe the data and method employed to examine the orientation of galaxy spin with the cosmic web.
\subsection{Observational samples and galaxy spin vectors}
Our study is based on galaxy positions in the 2MASS Redshift Survey~\citep[2MRS,][]{2mrs}  data sample.  The 2MRS is an all sky infra-red spectroscopic survey of galaxies in the nearby universe 
out to a distance of $200\,\mathrm{Mpc}$.  The 2MRS catalog consists of $44,599$ galaxies. The classification of 2MRS galaxies into elliptical and spiral galaxies has been adopted from \cite{2mrs}. 
We consider two subsamples of 2MRS: (1) all 2MRS galaxies whose distances are less than $115\, {\mathrm {Mpc}}$; (2) those 2MRS galaxies that are also in the Cosmicflows-2 catalog and hence 
used in the Wiener filter reconstruction of the local universe. We choose galaxies within specified distances for both samples, for reasons that will be made clear later.  The number of galaxies in each sample used in this work are presented in Table~1.
\begin{table}
\begin{center}
\caption{The number of galaxies in each sample used in this study. The sub-sample `2MRS $\cap$ CF2' 
is selected from the 2MRS sample but has only galaxies which are in CF2 and within $65\,h^{-1} \mathrm{Mpc}$.}
\begin{tabular}{p{1.9cm}@{} |p{1.8cm}|p{1.8cm}|p{1.8cm}|}
\hline
\hline
Sample  & $N_{\rm total}$ & $N_{\rm spirals}$ & $N_{\rm ellipticals}$  \\

\hline
\hline
2MRS & 19\,438& $11\,104$ & $8\,334$\\
2MRS $\cap$ CF2 & 2\,650 & $2\,026$ & $624$ \\
\hline
\end{tabular}
\end{center}
\label{table1}
\end{table}

This observational survey provides the galaxy's position angle and the isophotal axis ratios. Along with the inclination angle, these are the ingredients required to compute each galaxy's projected short axis. We assume the {\it projected} short axis and the spin axis are parallel. In principle, there can be misalignment between the shape of the system and its angular momentum vector. There are two reasons why the short axis may not be parallel to the spin axis. The system may simply not be rotationally but rather anisotropically supported. Or the two vectors may not align because we are observing the systems in projection (the orientation of a triaxial galaxy towards an observer is random).  Interestingly, analysis presented by \cite{franx+1991}  showed that majority of early-type galaxies indeed had small mis-alignments \citep[See also,][]{krajnovi+2011}. The most modern estimates on elliptical galaxy spin axes claim that a majority of elliptical galaxies appear like spiral galaxies with the gas and dust removed \citep{Cappellarim+2011}, namely with spin vectors aligned with their short axes. Thus, in this work, we assume that direction of the spin axis of an elliptical galaxy is given by its principal axis, fully aware that this is a simplifying approximation.

Since the data tells us the {\it projected} spin axis, and we require the {\it 3D} spin axis, we need to somehow break the line-of-sight degeneracy regarding the galaxy's inclination. There are a variety of ways suggested in the literature from limiting one's sample only to edge-on galaxies \citep[e.g.][]{navarro2004, jones+2010} to applying some kind of galaxy modeling \citep{lee+2007}.

For our purpose, we use the method as described in \citep{lee+2007} to determine the inclination angle. However, \cite{lee+2007} analysed a sample of spiral galaxies. Assuming that the {\it projected} short axis and the spin axis are parallel, one can extend the same method for elliptical galaxies. This method involves assuming a specific galaxy type has an intrinsic flatness parameter which, when combined with the isophotal axis ratio, can return an inclination angle \citep[See also,][]{haynes+1984}. Since it is not possible to determine the sign of the inclination angle, we have two spin vectors for each galaxy, one tending towards the observer and one tending away (in projection). As we have no way of knowing which is the correct inclination angle for a given galaxy, we simply consider both.  The effect of this is surely to weaken any intrinsic signal we find. Our results can thus be interpreted as a lower limit. Each spin vector is normalized to unity. Other authors chose to use just one ~\citep{lee+2007} or both~\citep{kashikawa+1992,varela+2012} spin vector simultaneously (as we do) or use some other approach to deal with this indeterminacy of the spin vectors of galaxies~\citep{trujillo+2006,slosar+2009}. 
\subsection{Cosmicflows-2 and the Wiener Filter Velocity field reconstruction}
We use the Cosmicflows-$2$ \citep[CF$2$,][]{tully+2013} catalog of distances as a starting point for our Wiener Filter 
reconstruction ~\citep{courtois+2012, courtois+2013}. The $\sim8000$ galaxy distances in CF2 are gathered according to the cosmic distance ladder and are thus from an amalgum of six methodologies: Cepheid variables, Tip of the Red Giant Branch, Surface brightness fluctuations, Tully-Fisher relation, Fundamental plane and SuperNovae Ia. The coverage is dense and accurate locally, becoming sparse beyond $6\,000 \,\mathrm{km} \,\mathrm{s}^{-1} \,(\sim86 \, {\mathrm {Mpc}})$and very sparse at $20\,000\, {\rm km}\, {\rm s}^{-1} \,(\sim285 \, \mathrm{Mpc})$. Peculiar velocities are decoupled from the Hubble flow, given an independent distance measure and the redshift. We choose the threshold of $8000\,  {\rm km}\, {\rm s}^{-1}$ as a balance between the sparse coverage and the volume available which corresponds to distance of $\sim 115\,\mathrm{Mpc}$. This makes the distance-limit to our first sample.

A Wiener filter (WF) is applied to CF2 to obtain the full 3D density and velocity fields \citep[see][]{zaroubi+1999, hoffman+2015}. The WF reconstruction is  essentially a minimum variance estimate which, given a prior (in this case a gaussian random field of perturbations according to the $\Lambda$CDM cosmological model) returns the statistically most likely velocity and density field. We assume standard WMAP7 \citep{komatsu+2011} cosmological parameters:   
(i.e. $\Omega_0 \,=\, 0.279, \, \Omega_b\, =\,0.046, \, \Omega_{\Lambda}\,=\, 1-\Omega_0, \,H_0\,=\,100h\,=\,70~\mathrm{km}~\mathrm{s}^{-1}\mathrm{Mpc}^{-1}, \, \sigma_8=0.817$).

The accuracy of the WF is limited by the sampling and the measurement error of the constraints. The more peculiar velocities we have accurately measured, the better the accuracy of the WF. In the case of CF2, the spatial resolution of the WF is around $2.5\,h^{-1} \mathrm{Mpc}$. Small scale power below this limit can be added randomly from the power spectrum in the form of Constrained Realizations (CRs). A set of $20$ CRs is used to estimate the robustness of the WF.

\subsection{Shear Tensor}
Once the full 3D velocity field is known from the WF reconstruction, the principal axes of the cosmic shear field  can be obtained. The shear tensor is defined as the (normalized) symmetric part of the deformation of the velocity field: 
\begin{equation}
\Sigma_{\alpha\beta}=-\frac{1}{2H_{0}}\bigg(\frac{\partial v_{\alpha}}{\partial r_{\beta}}+\frac{\partial v_{\beta}}{\partial r_{\alpha}}\bigg),
\end{equation}
where $\alpha$ and $\beta$ are the $x, y$, and $z$ components of the velocity $v$ and position $r$, and $H_0$ is Hubble's constant. The negative sign is added for convention.  The shear tensor is diagonalized and the eigenvalues ($\lambda_1, \, \lambda_2,$ and $\lambda_3$) and corresponding eigenvectors (\eone, \etwo and \ethree) are then obtained.  Eigenvalues are then ordered as $\lambda_1>\lambda_2>\lambda_3$, denoting the strength of compression or expansion along the eigenvectors~\citep{hoffman+2012,Libeskind+2012,noam+2013}. Note that the shear tensor's  eigenvectors are non-directional lines, and as such their direction has a degeneracy of 180 degrees.
\begin{table*}
\begin{center}
\begin{tabular}{|l|c|c|c|c|c|c|}
\hline
 &\multicolumn{2}{| c|} { $|\hat{\rm L} \cdot \hat{\rm e}_1| $} &\multicolumn{2}{|c|} { $|\hat{\rm L} \cdot \hat{\rm e}_2| $}&\multicolumn{2}{|c|} {$|\hat{\rm L} \cdot \hat{\rm e}_3| $}\\
\hline
\hline
&\multicolumn{2}{| c|} {} &\multicolumn{2}{|c|} {}&\multicolumn{2}{|c|} {}\\
{\bf Median cosines} &\multicolumn{2}{|c|} {} &\multicolumn{2}{|c|} {}&\multicolumn{2}{|c|} {}\\
Spirals \hspace*{0.46cm} 2MRS&\multicolumn{2}{|c|} {$0.49$} &\multicolumn{2}{|c|} {$0.51$}&\multicolumn{2}{|c|} {$0.51$}\\
Spirals \hspace*{0.46cm} 2MRS$\cap$CF2 &\multicolumn{2}{|c|} {$0.46$} &\multicolumn{2}{|c|} {$0.52$}&\multicolumn{2}{|c|} {$0.51$}\\
Ellipticals \hspace*{0.1cm} 2MRS&\multicolumn{2}{|c|} {$0.57$} &\multicolumn{2}{|c|} {$0.48$}&\multicolumn{2}{|c|} {$0.47$}\\
Ellipticals \hspace*{0.1cm} 2MRS$\cap$CF2 &\multicolumn{2}{|c|} {$0.60$} &\multicolumn{2}{|c|} {$0.46$}&\multicolumn{2}{|c|} {$0.46$}\\
&\multicolumn{2}{| c|} {} &\multicolumn{2}{|c|} {}&\multicolumn{2}{|c|} {}\\
\hline
& $p^{\rm mean}_{\rm KS}$ &MP&$p^{\rm mean}_{\rm KS}$&MP&$p^{\rm mean}_{\rm KS}$&MP\\
\hline
 &  &&&&&\\
{\bf galaxy and eigenframe} &&&&&&\\
{\bf randomization} &  &&&&&\\
Spirals \hspace*{0.46cm} 2MRS&$  5.5\times10^{-4 }$ &$0.3$&$  2.0\times10^{-1 }$&$86.5$&$  1.9\times10^{-3}$&$96.8$\\
Spirals \hspace*{0.46cm} 2MRS$\cap$CF2 & $1.4\times10^{-5 }$ &$0.0$&$  3.8\times10^{-3 }$&$99.3$&$  1.2\times10^{-1}$&$93.3$\\
Ellipticals \hspace*{0.1cm} 2MRS & $  2.9\times10^{-34 }$ &$100.0$&$  2.4\times10^{-3 }$&$0.2$&$  2.5\times10^{-14}$&$0.0$\\
Ellipticals \hspace*{0.1cm} 2MRS$\cap$CF2 & $  1.3\times10^{-10 }$ &$100.0$&$5.0\times10^{-3 }$&$4.2$&$  1.8\times10^{-4}$&$0.0$\\
 &  &&&&&\\
\hline
 &  &&&&&\\
{\bf CR randomization} &&&&&&\\
Spirals \hspace*{0.46cm} 2MRS&$2.98\times10^{-1}$  &$4.761$&$3.57\times10^{-1}$&$80.952$&$5.66\times10^{-1}$&$47.619$\\
Spirals \hspace*{0.46cm} 2MRS$\cap$CF2 & $1.45\times10^{-1}$ &$4.761$&$1.42\times10^{-1}$&$90.476$&$5.25\times10^{-1}$&$33.333$\\
Ellipticals \hspace*{0.1cm} 2MRS & $3.54\times10^{-2}$ &$95.236$&$3.36\times10^{-2}$&$4.761$&$4.35\times10^{-1}$&$42.857$\\
Ellipticals \hspace*{0.1cm} 2MRS$\cap$CF2 & $2.40\times10^{-1} $&$95.238$&$1.58\times10^{-1}$&$4.762$&$5.97\times10^{-1}$&$23.810$\\
 &  &&&&&\\
\hline
 &  &&&&&\\
\end{tabular}
 \caption{The top rows ({\bf ``Median cosines''}) show the median (cosine of the) angle formed between the eigenframe and $\hat{\mathrm{L}}$.  This indicates the alignment of the measured signal. The rest of the table displays the two statistical tests we have performed measuring $p^{\rm mean}_{\rm KS}$ and where the measured median cosine sits as a percentile amongst the 21  or 1000 trials - the median percentile (MP). The second part of the table ({\bf ``galaxy and eigenframe randomization''}) shows that the galaxies and eigenframes themselves do not bias one towards the signal. The third part of the table ({\bf ``CR randomization''}) shows that the WF is statistically fully consistent with the CRs.}
 \label{table1}
\end{center}
\end{table*}

\subsection{Estimating statistical significance of the alignment signal}
\label{sec:statsig}
We will examine the correlation of galaxy spin vectors with the principal axis of the shear field by computing  the probability 
distribution function $P(\mu)$, where $\mu$ is the cosine of the angle between the eigenvectors of the shear tensor computed at the location of each galaxy and the rotation axis of the galaxy, $\hat{\rm L}$. The quantity $\mu$ is determined as the dot product between the respective normalized vectors, namely $\mu \, =\,|{\rm cos}\, \theta| \, \in \, [0,1]$. $\mu\, = \,0$ implies that the rotation axis of the galaxy is perpendicular to the respective eigenvector and $\mu\,=\,1$ implies the two are
parallel. We use kernel density estimation technique~\citep[see appendix of][]{tempel+2014} to obtain a histogram of the probability distribution.

To obtain the statistical significance of any alignment signal, we need to compare the measured probability distribution function with the null hypothesis of a random alignment between eigenvectors and galaxies. In an ideal world, a random alignment would correspond to a uniform distribution of $P(\mu)=1$.  However, due to selection effects, neither the galaxies nor the reconstructed peculiar velocity field have a uniform distribution with the line-of-sight. \cite{tempel+2013} examined the distribution of the inclination angles of galaxies in observations and found an excess of edge on spirals, implying that many face-on spiral galaxies are missing from the data set. This is clearly  a selection effect. To overcome this kind of selection effect, we use a  Monte-Carlo method similar to that used in~\cite{tempel+2013} to compute the probability distribution for a random orientation between galaxy spin vectors and shear eigenvectors. Our starting point is the distribution of angles between the galaxies' spin vectors and the line of sight. We re-assign each galaxy a random orientation drawn from the full sample\footnote{In practice a random galaxy's position angle is combined with another random galaxy's inclination angle to compute a new randomized spin vector. This double-randomization ensures that the distribution of position and inclination angles does not change, thereby preserving any inherent biases in these quantities.}.
We then randomly select an eigenframe (without repetition) that is assigned to another galaxy in the same morphology/sample. Thus for ellipticals in the 2MRS, we select randomly from the eigenframes that are assigned to the 2MRS elliptical galaxies. Same for spirals in 2MRS, etc. We do this $1000$ times, each time computing the angle formed between the (randomized) spin vector and the (randomized) shear eigenframe. For each of the $1000$ trials, we obtain a probability distribution.
The median, $3$\%, and $97$\% percentiles from these randomized probability distributions is used to estimate the statistical significance of the measured signal. 
This method of finding the random probability distribution takes care of the biases in the observations of the galaxies and the reconstruction of the peculiar velocity field. However, the randomized probability distribution is not uniform ($\sim1$) due to the line of sight bias. 
The region spanned by shuffling galaxies and eigenframes in this way is shaded as `blue' in the figures and tells us what we would expect with the full scatter of a random distribution. If our signal lies within the randomized one, there is $100$\% chance that it is consistent with random.  We characterize the strength of our measured signals by performing a Kolmogorov-Smirnov (KS) test which measures a probability, $p_{{\rm KS}}$ that the measured set of angles and the randomized one, are drawn from the same parent. These are presented in Table 2 and the result of such a shuffling between galaxies and eigenframes can be seen in the second part titled {\bf ``galaxy and eigenframe randomization''}.

The KS test is one way to test whether two signals are consistent with eachother.
Another method is to compare the median of the measured alignment distribution to that of random distribution.
For each of the $1000$ randomized trials described above, we compute the median, $\alpha$. The ensemble of $\alpha$'s provide us with a feeling for what kind of median to expect from randomized distributions. We then calculate the median percentile (MP) of the measured signal from these randomized distributions. If MP is 0.01, then the chance of getting such a measured signal is 10 out of 1000 random draws. We perform this analysis for all figures in this paper and check the measured strength of the signal in this way.   Table 2 also lists the median value of the cosine of the angle formed between the spin vector and shear eigenvectors of the measured signal under the name {\bf Median cosines}. 

There is one more additional bias due to the uncertainty of the WF reconstruction, since the WF reconstruction is a result of sparse sampling. In order to get a feeling for how robust the eigenvectors of the WF are, we make use of $20$ constrained realizations (CRs). The CRs converge to the WF on large scales, but small scale power is added randomly, where the WF has none. Note that a large set of CRs represent the most conservative estimate of the reconstructed velocity (and shear) field. On scales where the WF is robust, the added small scale power of the CRs have no affect and therefore the CR and  WF eigenvectors are all well aligned. In regions where there is no data and where the WF estimation thus converges to the mean field, there can be a misalignment in the directions of the CR and WF eigenvectors. The galaxy-spin shear field signal is thus conservatively estimated by each of the CRs and the range obtained is shaded as `grey' in the last two figures.  For each CR signal, we also conduct a KS test between it and the WF signal to test the hypothesis that the two have the same parent. The KS test returns a probability $p_{\rm KS}$ that this hypothesis is not ruled out: large values of $p_{\rm KS}$ indicate that the distributions do not differ considerably. The KS tests and Median Percentiles (MP)  are summarized in the third part of the Table 2 (under the heading {\bf CR randomization}). 

\section{Results}
\label{sec:results}

\subsection{Spin alignment in 2MASS redshift survey data}
 \begin{figure*}
\includegraphics[width=0.95\textwidth]{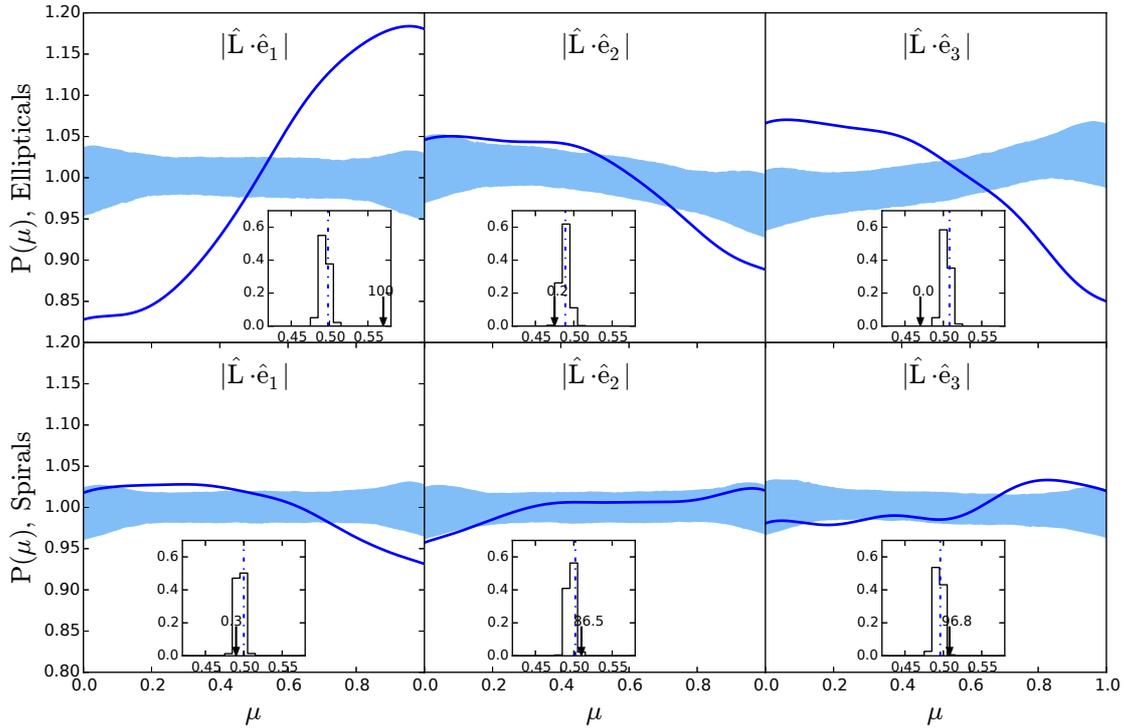}
 \caption{The probability distribution P($\mu$) as a function of the cosine of angle formed, $\mu\, = \,|{\rm cos} \, \theta|$, between the galaxy 
 spin vector and the cosmic web for the first 2MRS sample. The upper and lower panels correspond to the elliptical and spiral galaxies respectively whereas the left, middle and the right show the alignments correspond to $\mu \, = \, |\hat{\rm L} \cdot \hat{\rm e}_1|, \, |\hat{\rm L} \cdot \hat{\rm e}_2|, \, |\hat{\rm L} \cdot \hat{\rm e}_3|$ respectively.  The blue shaded region shows the $3$th and $97$th percentile spread of a randomised distribution and the blue solid line shows the observed alignment signal.  The inset of each panel shows the distribution of medians for 1000 randomised trials. The $x$-axis in the inset represents the median $\mu$'s and the $y$-axis shows the fraction of the number of trials. The dotted blue line shows the average of the medians for these randomised trials. The arrow denotes the value of the WF median and its percentile - with respect to the distribution of the other medians, is given as the number above the arrow. As can be inferred, the WF is statistically inconsistent with the randomised trials.}
 \label{fig:2mrs}
 \end{figure*}

In this section, we examine the orientations of spin vectors of 2MRS galaxies with respect to the eigenvectors of the shear field. We start by dividing first sample into elliptical and spiral galaxies. Fig.~\ref{fig:2mrs} shows the probability distribution of $\mu=|\cos\theta|=|\hat{{\rm L}}\cdot \hat{\rm e}_{i}|$ for elliptical (left column) and spiral (right column) galaxies for \eone (top), \etwo (middle) and \ethree (bottom).  Elliptical galaxies  spin parallel to \eone and perpendicular to \ethree. KS tests presented in the third part of Table 2 confirm this: the KS probability that the alignment with \eone and \ethree, seen for elliptical galaxies is drawn from random is $ 2.9\times10^{-34 }$ and $  2.5\times10^{-14}$. The same impression is gleaned when examining how likely the measured median percentile (MP) is - if it were drawn from a large set of random probability distributions. This is shown as an inset in the same figure. Here we show the distribution of medians for each of the 1000 random trials, along with an arrow that indicates the median of the WF reconstruction. Clearly the WF median is well outside of the range produced by the randomization process: the median alignment with \eone and \ethree correspond to the $100$ and $0.0$ percentiles, respectively (namely are fully outside the range expected from random). This shows that measured alignment of elliptical galaxies is very strong with the \eone and \ethree eigenvectors. For spiral galaxies, the probability distributions are consistent with random as gleaned from the shapes of distributions in Fig.~\ref{fig:2mrs}.

There can be three reasons that no signal is seen for spiral galaxies. Perhaps there is no spin alignment of spiral galaxies on the scales testable by CF2 reconstruction. Perhaps there is a signal, but its too weak to be statistically isolated and possibly, the degeneracy of the inclination angle wipes it out. Given the large sample size of 2MRS dataset, this seems unlikely.  Alternatively, the eigenvectors of the shear tensor could be at fault.  Perhaps the 2MRS data extends into regions where the WF reconstruction is poor, regions which are barely covered by CF2. To test this hypothesis, we examine the alignment signal just for those 2MRS galaxies that are also in CF2 and within distance of $65\, h^{-1} \mathrm{Mpc}$. This sample is denoted as `2MRS $\cap$ CF2'. By doing so,  we specifically bias ourselves just to those galaxies that are used as the best constraints in the WF reconstruction.  Beyond $65\, h^{-1} \mathrm{Mpc}$, the distance errors in these galaxies are large enough to make the WF reconstruction inaccurate. Hence, the shear field at these close distance should be the best recovered one. In other words, if no alignment is seen for even these galaxies - where the WF is best defined and the velocity field is best constrained - then probably there is no alignment of spiral galaxies in the first 2MRS sample.

\begin{figure*}
\includegraphics[width=0.95\textwidth]{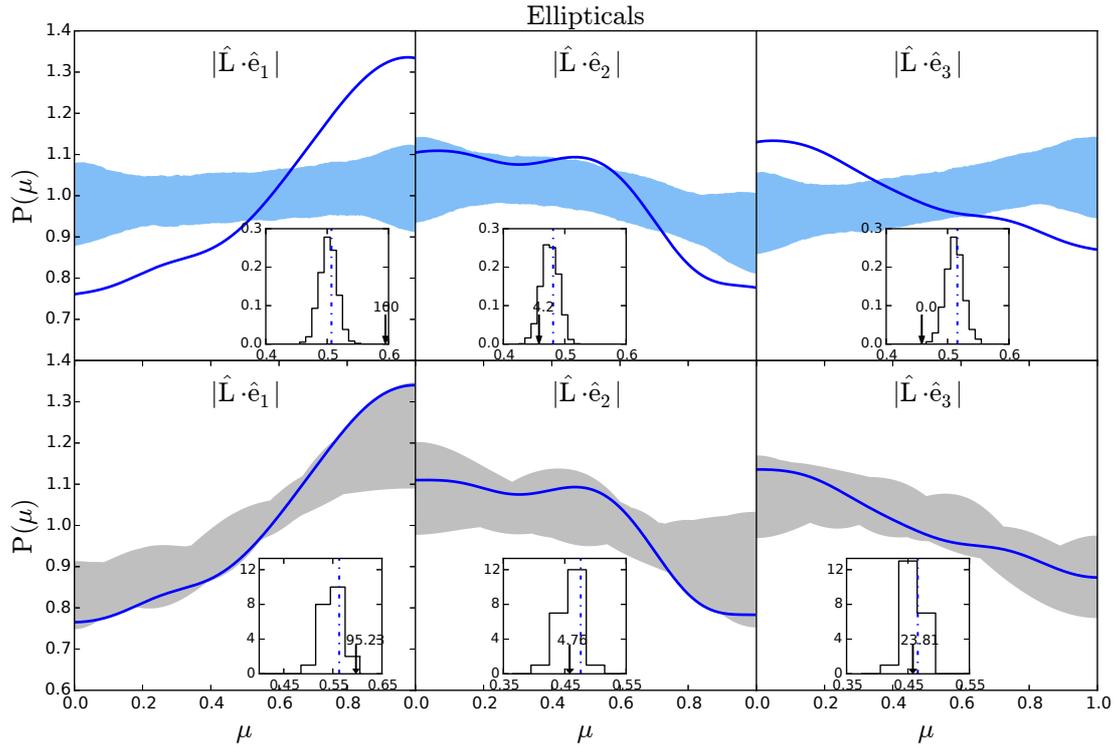}
\caption{ The probability distribution P($\mu$) as a function of the cosine of angle
formed, $\mu\, = \,|{\rm cos} \, \theta|$, between the galaxy spin vector and the cosmic web for the elliptical galaxies in the 2MRS $\cap$ CF2 sub-sample.  The left, middle and the right show the alignments correspond to $\mu \, = \,|\hat{\rm L} \cdot \hat{\rm e}_1|, \, |\hat{\rm L} \cdot \hat{\rm e}_2|, \, |\hat{\rm L} \cdot \hat{\rm e}_3|$ respectively.  The blue shaded region (upper panel) shows the $3$th and $97$th percentile spread of a randomised distribution and the blue solid line shows the observed alignment signal whereas the grey shaded region (lower panel) indicates the scatter on the signal due to the $21$ Constrained realizations (including WF).  The inset of upper panel shows the distribution of medians for 1000 randomised trials whereas the inset of lower panel shows the distributions of medians for 21 CRs. The $x$-axis in the inset represents the median $\mu$'s and the $y$-axis shows the fraction of the number of trials (upper inset panel) or the number of CRs (lower inset panels).  The dotted blue line shows the average of the medians for these randomised trials. The arrow denotes the value of the WF median and its percentile - with respect to the distribution of the other medians, is given as the number above the arrow. As can be inferred, the WF is statistically inconsistent with the randomised trials.}
 \label{fig:2mrscf2_ellip}
 \end{figure*}

  \begin{figure*}
 \includegraphics[width=0.95\textwidth]{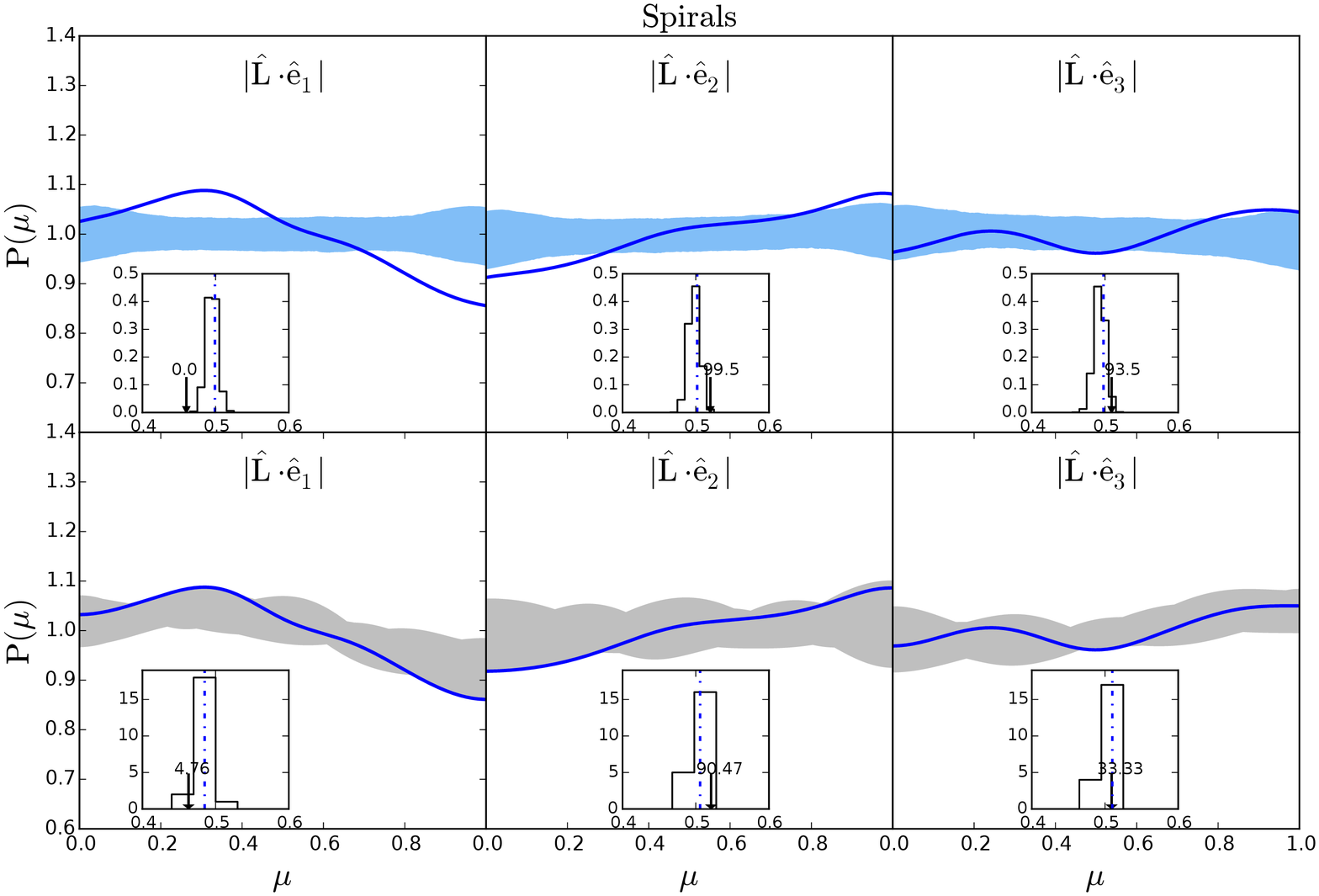}
 \caption{The probability distribution P($\mu$) as a function of the cosine of angle
formed, $\mu\, = \,|{\rm cos} \, \theta|$, between the galaxy spin vector and the cosmic web for the spiral galaxies in the  2MRS $\cap$ CF2 sub-sample.  The left, middle and the right show the alignments  correspond to $\mu \, = \, |\hat{\rm L} \cdot \hat{\rm e}_1|, \, |\hat{\rm L} \cdot \hat{\rm e}_2|, \, |\hat{\rm L} \cdot \hat{\rm e}_3|$ respectively.  The blue shaded region (upper panel) shows the $3$th and $97$th percentile spread of a randomised distribution and the blue solid line shows the observed alignment signal whereas the grey shaded region (lower panel) indicates the scatter on the signal due to the $21$ Constrained realizations (including WF). The inset of upper panel shows the distribution of medians for 1000 randomised trials whereas the inset of lower panel shows the distributions of medians for 21 CRs. The $x$-axis in the inset represents the median $\mu$'s and the $y$-axis shows the fraction of the number of trials (upper inset panel) or the number of CRs (lower inset panels).  The dotted blue line shows the average of the medians for these randomised trials.  The arrow denotes the value of the WF median and its percentile - with respect to the distribution of the other medians, is given as the number above the arrow. As can be inferred, the WF is statistically inconsistent with the randomised trials.}
 \label{fig:2mrscf2_spir}
 \end{figure*}

In Fig.~\ref{fig:2mrscf2_ellip} and~\ref{fig:2mrscf2_spir}, we show the probability distribution for elliptical and spiral galaxies in the 2MRS $\cap$ CF2 sample respectively. The blue shaded region (top row) shows the $3$\% and $97$\% percentile spread of the randomized distribution whereas the grey shaded region (bottom row) gives the range expected from the spread in the constrained realizations. In each inset panel, a histogram of the medians from either sample is indicated along with an arrow for the measured median. The upper panels of Fig.~\ref{fig:2mrscf2_ellip} show that elliptical galaxies are aligned parallel to \eone and perpendicular to \ethree. The strength of the signal is stronger for this sample in comparison to 2MRS as depicted in figures (e.g see the $y$-axis range). This increase is expected since, as mentioned above, this region has a more data and thus more accurate reconstructed velocity field. The KS probability that the alignment for $|\hat{\rm L}\cdot \hat{\rm e}_{1}|$ and $|\hat{\rm L}\cdot \hat{\rm e}_{3}|$  are statistically significant, is shown in Table 2 and is estimated at $1.3\times10^{-10 }$ and $  1.8\times10^{-4 }$. Similarly, the median of their distributions are at $100$ and $0$ percentiles.

The lower row of Fig.~\ref{fig:2mrscf2_ellip} confirms that the reconstruction is stable for this sample as the signal is within the scatter of the CRs.  This is our main result: {\it elliptical galaxies show a statistically significant tendency to be aligned with the eigenvectors of the cosmic shear field.}  Note that no alignment is found with \etwo, just with \eone and \ethree.

Note that the strongest alignment found here is for those galaxies with the poorest defined spin axis --ellipticals. \cite{tempel+2013} and \cite{tempel+2013b} have also found that alignment is stronger for ellipticals than spirals. It is thus important to conclude here: {\it the intrinsic alignment for ellipticals is much stronger than the alignment for spiral galaxies}. An alignment between the spin axis of spirals and \eone is also seen in Table.~2 although this is not completely obvious from the probability distribution in Fig.~\ref{fig:2mrscf2_spir}. The KS test indicates that it is highly unlikely to be drawn from a random distribution; this is echoed by the fact that the measured median is at 0 percentile when compared with random trials. No (statistically significant) alignment is seen for either of the two other eigenvectors (\etwo or \ethree).

Such a weak alignment for spirals is consistent with that found by \cite{tempel+2013,tempel+2013b}, although  those studies only examined filaments, whereas our study presents a more elaborate picture on the aspect of web environment in the sense that the shear is more general: the  2MRS $\cap$ CF2 is not prejudiced towards one kind of environment over another. In other words, we find the alignment of galaxies not just with filaments (the \ethree direction) but also with sheets and voids. Since previous studies have reported finding correlations between galaxy spin  and filament axes, some alignment with the shear tensor is expected here. 

  \subsection{Spin alignment in the  Local Group}
The shear eigenvectors at the location of the Local Group have been published by \cite{libeskind+2015}. These suggest that the Local Group (LG) resides in a filament compressed by the expansion of the Local Void and stretched by the Virgo cluster. In this section, we examine how the Local  Group spin is aligned with the shear. The spin axis of the Milky Way ($\hat{\rm L}_{\rm MW}$) is often given as  (RA, Dec = $12^{\rm h} 51.4^{\rm m}, \,+27^{\circ}07'$) while if Andromeda's (M31's) position angle is known and (RA, Dec = $00^{\rm h}41.8^{\rm m},\, +41^{\circ}16'$),  its spin axis ($\hat{\rm L}_{\rm M31}$) can be easily computed. Computing the angle formed by these  spin axes with the \eone published by \cite{libeskind+2015}, leads to $|{\rm cos}(\hat{\rm e}_1\cdot\hat{\rm L}_{\rm MW})|\, =\, 0.22$ and $|{\rm cos}(\hat{{\rm e}}_1\cdot\hat{\rm L}_{\rm M31})| \, = \, 0.63$.  Just as in the 2MRS sample, these two spiral galaxies show no striking alignment with the shear. 

We now turn to the orbital angular momentum of the Local Group, $\hat{\rm L}_{\rm LG}= r_{\rm M31} \times (v_{\rm rad}\hat{r}+v_{\rm tan}\hat{\theta})$ and how this aligns with the cosmic shear field. $v_{\rm rad}$ has been well established since Vesto Slipher first measured M31's blueshift a century ago~\citep{silpher1913}. \cite{marel2012} have measured the proper motion of M31 finding that $v_{\rm tan} \approx 17\pm17$ km/s. Since this painstaking measurement is notoriously difficult to perform, the 1$\sigma$ confidence region of $v_{\rm tan}$ is also consistent with zero. Of course, if $v_{\rm tan}=0$ and the orbit is purely radial, then $\hat{\rm L}_{\rm LG} =0$ since the spin axis is undefined and the angular momentum is null.

If however, we assume the directions of $v_{\rm tan}$ measured by \cite{marel2012} \citep[and presented in SG coordinates in Table 3  by][]{shaya+2013}, we can compute a well defined direction for $\hat{\rm L}_{\rm LG}$ and thus a well defined $|\cos(\hat{\rm e}_{\rm 1}\cdot\hat{\rm L}_{\rm LG})|$. For the favored value of the direction of $v_{\rm tan}$, this lies roughly in (\etwo-\ethree)  plane: $|\cos(\hat{{\rm e}}_{\rm 1}\cdot\hat{\rm L}_{\rm LG})| =0.95$. Within the experimental uncertainty, the direction of $v_{\rm tan}$ may differ by up to $\sim$55 degrees from the favored value, a direct result of the small value of $v_{\rm tan}$ \citep[if $v_{\rm tan}$ were greater, the error on its direction would be much smaller, see][]{pawlowski+2013}. Of course given such a large experimental error on $v_{\rm tan}$ and $\hat{\rm L}_{\rm LG}$, the alignment with \eone should be viewed with caution.

Nevertheless, assuming the angular momentum of the M31-Milky Way merger is conserved, the final (elliptical) remnant of the collision will likely be spin aligned with the \eone direction, following the trend seen for elliptical galaxies in the 2MRS. Note that this is exactly the same as the prediction made by \cite{foreroromero+2015} who examined a suite of Local Group Like pairs in a large numerical simulation. In a word, our Local Group is a proto-type of the alignments measured in the 2MRS: as individual spiral galaxies no convincing alignment is seen, but as an elliptical merger remnant, alignment with the shear field is forthcoming.

 \section{Summary and Discussion}
 \label{sec:concl}

The accepted theory of how galaxies obtain their spin is from torques imparted in the linear regime due to the misalignment of a proto-halo's inertia tensor and the tidal field. This mechanism should leave its imprint on the alignment of galaxies with the tidal field. Specifically TTT predicts that spin should be preferentially aligned along the intermediate axis of the tidal field. A number of observational studies \citep{kashikawa+1992, navarro2004, trujillo+2006,tempel+2013,tempel+2013b}
regarding the spin alignments of the galaxies with features in the tidal field have been carried out. Very few have found the alignment predicted by TTT. Indeed not all were looking specifically for an alignment with the intermediate axis of the tidal field. \cite{lee+2007} did find a statistically significant alignment with the intermediate axis using a reconstruction  of the tidal field. To date, this remains one of the only studies to have done so.

Insights from simulations suggest spins align with the eigenvector corresponding to weakest collapse (\ethree) for small haloes and with the eigenvector corresponding to strongest collapse (\eone) for the most massive haloes \citep{Aragon-Calvo+2007,Codis+2012,noam+2013}, although the mechanism  that causes this spin flipping is still disputed \citep{trowland+2013}. For example, see also \cite{codis+2015}. \cite{Codis+2012} claims that this spin ``flip'' occurs because large haloes grow by mergers along  filaments (i.e. mergers along \ethree that are perpendicular to \eone), while small haloes accrete material along \eone. By using reconstructions  of the tidal field, we can test this spin flip in all environments, namely not just where filaments are well defined entities.

We have used 2Mass Redshift Survey galaxy catalog \citep{2mrs} to investigate the orientations of spin axes of galaxies with respect to the eigenvectors of the cosmic tidal shear tensor.   We assume that any galaxy's {\it projected} short axis is parallel to its spin axis and we note that ample work has been done related to this \citep{franx+1991,Cappellarim+2011}, showing that the majority of the early-type galaxies have small mis-alignments between their shapes and the angular momentum vectors. It is also consistent with the measured shape alignments of ellipticals and the tidal alignment model \citep{hirata+2004} such as  in the analysis conducted using the MegaZ-LRG sample \citep{joachimi+2011} and the SDSS data \citep{hirata+2007}. This has also been discussed in the recent reviews \citep{joachimi+2015,kirk+2015,kiessling+2015}. The shear field is determined by a Wiener filter reconstruction of the Cosmicflows-2 (CF2) survey of peculiar velocities.

In the 2MRS catalog (within $115\,\mathrm{Mpc}$), a statistically significant signal is seen (see Fig.~\ref{fig:2mrs}) indicating a correlation between the spin axes of elliptical galaxies and eigenvectors of shear tensor. This signal is significantly strengthened when we specifically chose just those galaxies that are in both 2MRS and CF2 and within $65\, h^{-1} \mathrm{Mpc}$, namely, we have examined just those galaxies where the reconstructed velocity field is the best constrained. It is impressive that with such a small sample ($\sim600$ galaxies) a strong statistical significant signal is seen (see Fig. 2), whereas in numerical simulations, order of magnitudes more haloes are needed to convincingly demonstrate a statistical signal. These elliptical galaxies tend to align perpendicular to the \ethree eigenvector and parallel to the  \eone eigenvector, a similar trend seen in numerical simulations. This is a very significant result which has been seen for the first time with the Large Scale structure reconstructed by the peculiar velocity field.

The reader will note that the results presented here are based entirely on the Wiener Filter (WF) reconstruction method. The WF tends to the mean field in regions with fewer or no constraints. Small scale power can be added at random in a method called constrained realizations (CRs). Since adding small scale velocity effects at random is the most drastic way of modeling regions without data, the CRs provide the most conservative conservative estimate of the alignment signal. We have shown that even when using the CRs the alignment trend is the same, and statistically inconsistent with random, although strength of the alignment is weaker when using the CRs (as expected.)

In this study, no convincing correlation  is found between the spin axes of spiral galaxies and the eigenvectors of the shear tensor.  A tendency is seen in the form of slight signal with \eone. Such correlations are found in the $N-$body simulations \citep{Aragon-Calvo+2007,noam+2013} and in one observational study \citep{lee+2007}. There are a number of reasons why we have not been able to measure such a strong signal. First, we probably need a larger spiral galaxy sample size. The alignment signal is weak, and given the limited number of galaxies, we may simply need a larger sample to see it. Secondly more data to properly reconstruct the peculiar velocity field and thus the shear field would improve the situation. It is possible that the combination of a weak signal to begin with is simply wiped out by reconstructions inaccuracies. More data, specifically in the form of a larger and deeper sky surveys combined with denser sampling of the peculiar velocity field will likely improve the situation.

\section*{Acknowledgments}
The authors thank the anonymous referee for important suggestions and comments that led to an improvement in the presentation of the paper.  IP acknowledges the support from the Leibniz-DAAD research fellowship. NIL is supported by the DFG. ET is supported by the grants IUT26-2, IUT40-2 of the Estonian Ministry of Education and Research. JS acknowledges support from the Alexander von Humboldt foundation.

\bibliography{allrefs}

\end{document}